\documentclass[useAMS,usenatbib]{mn2e}
\IfFileExists{srcltx.sty}{\usepackage[active]{srcltx}}

\usepackage{amssymb}
\usepackage{amsmath}
\usepackage{txfonts}
\usepackage{graphicx}
\usepackage{natbib}
\usepackage{rotating}
\usepackage{url}
\usepackage{varioref}
 
\newcommand{\kev}{\:\mathrm{keV}} 

\begin{document}

\title{Constraints on sterile neutrino as a dark matter candidate from the
  diffuse X-ray background} %

\author[Boyarsky et al.]{A.~Boyarsky$^{1,2,3}$, A.~Neronov$^{4,5}$,
  O.~Ruchayskiy$^{6}$, M.~Shaposhnikov$^{2,1}$
  \\
  $^{1}$CERN, Theory department, Ch-1211 Geneve 23,
  Switzerland\\
  $^{2}$\'Ecole Polytechnique F\'ed\'erale de Lausanne, Institute of
  Theoretical Physics, FSB/ITP/LPPC, BSP 720, CH-1015, Lausanne,
  Switzerland\\
  $^{3}$On leave of absence from Bogolyubov Institute of Theoretical Physics,
  Kyiv, Ukraine\\
  $^{4}$INTEGRAL Science Data Center, Chemin d'\'Ecogia 16,
  1290 Versoix, Switzerland\\
  $^{5}$Geneva Observatory, 51 ch. des Maillettes,
  CH-1290 Sauverny, Switzerland \\
  $^{6}$Institut des
  Hautes \'Etudes Scientifiques, Bures-sur-Yvette, F-91440, France\\
}

\date{Received $<$date$>$  ; in original form  $<$date$>$ }
\pagerange{\pageref{firstpage}--\pageref{lastpage}} \pubyear{2005}

\maketitle
\label{firstpage}

\begin{abstract}
  Sterile neutrinos with masses in the keV range are viable
  candidates for the warm dark matter. We analyze existing data for
  the extragalactic diffuse X-ray background for signatures of
  sterile neutrino decay. The absence of detectable signal within
  current uncertainties of background measurements puts
  model-independent constraints on allowed values of sterile neutrino
  mass and mixing angle, which we present in this work.
\end{abstract}

\section{Introduction}
\label{sec:intro}

At present time there exists an extensive body of evidence that most of the
matter in the Universe is composed of new, yet undiscovered particles -- dark
matter (DM).  Observations of (i) galactic rotation curves, (ii) cosmic
microwave background radiation, (iii) gravitational lensing, and (iv) X-ray
emission of hot gas in galaxy clusters provide independent measurements of DM
content of the Universe.

Another major experimental discovery of the recent decade is that of
neutrino oscillations.  There are separate measurements of neutrino
oscillations in solar neutrinos (\cite{SNO}), atmospheric neutrinos
(\cite{superK}), and reactor neutrinos (\cite{Kamland}).  Neutrino
oscillations can be explained if neutrino is a massive particle,
contrary to the Standard Model assumption. This means that along with
the usual left-handed (or \emph{active}) neutrinos there may exist
also right-handed or \emph{sterile neutrinos}.

Conventional sea-saw mechanism
(\cite{Minkowski,Yanagida,gell,ramond,mohapatra,glashow}) of generation of
small active neutrino masses implies that the sterile neutrinos are heavy
(usually of the order of GUT energy scale $\sim 10^{10}- 10^{15}$~GeV) and
that their mixing with usual matter is of the order $\sin\theta \sim 10^{-10}
- 10^{-15}$. In addition of the smallness of neutrino masses, models of this
type can explain baryon asymmetry of the Universe via thermal leptogenesis
(\cite{fukugita}) and anomalous electroweak fermion number non-conservation
(\cite{kuzmin}). However they do not offer a DM candidate.

Recently it was proposed that neutrino oscillations, the origin of
the dark matter, and baryon asymmetry of the Universe can be
consistently explained in the model called \emph{neutrino Minimal
Standard Model} ($\nu$MSM)~(\cite{Misha05a,Misha05b}). This model is
a natural extension of minimal Standard Model (MSM), where three
right-handed neutrinos are introduced into the MSM Lagrangian. In
this extension neutrinos obtain Dirac masses via Yukawa coupling
analogous to the other quarks and leptons of MSM and in addition
Majorana mass terms are allowed for right-handed neutrinos. Unlike
conventional see-saw scenarios, all of these Majorana masses (which
are roughly equal to the masses of corresponding sterile neutrinos)
are chosen such that the mass of the lightest sterile neutrino is in
the keV range and the other ones are $\lesssim 100$~GeV --- below
electroweak symmetry breaking scale. In this model a role of the dark
matter particle is played by the lightest sterile neutrino.  

The existence of a relatively light sterile neutrino has nontrivial observable
consequences for cosmology and astrophysics. It was proposed in~\cite{Dodelson:93}
that a sterile neutrino with the mass in the keV range may be a viable
``warm'' DM candidate. The small mixing angle ($\sin \theta \sim 10^{-6} -
10^{-4}$) between sterile and active neutrino ensures that sterile neutrinos
were never in thermal equilibrium in the early Universe and thus allows their
abundance to be smaller than the equilibrium one.  Moreover, a sterile
neutrino with these parameters is important for the physics of supernova
(\cite{fryer}) and was proposed as an explanation of the pulsar kick
velocities~(\cite{kusenko,2003PhRvD..68j3002F,2004PhRvD..70d3005B}).

In addition to the dominant decay mode into three active neutrinos, the light
(with mass $m_s\lesssim 1$~MeV) sterile neutrino can decay into an active one
and a photon with the energy $E_\gamma=m_s/2$.  Thus, there exists a
possibility of direct detection of neutrino decay emission line from the
sources with big concentration of DM, e.g. from the galaxy
clusters~(\cite{Fuller:01b}).  Similarly, the signal from radiative sterile
neutrino decays accumulated over the history of the Universe could be seen as
a feature in the diffuse extragalactic background light spectrum. This opens
up a possibility to study the physics beyond the Standard Model using
astrophysical observations.

Recently there has been a number of works devoted to the analysis of the
possibility to discover sterile neutrino radiative decays from X-ray
observations~(\cite{Fuller:01b,Mapelli:05}). For example, it was argued by
\cite{Fuller:01b} that if sterile neutrinos composed 100\% of all the DM, one
should be able to detect the DM decay line against the background of the X-ray
emission from the Virgo cluster. According to \cite{Fuller:01b} the
non-detection of the line puts an upper limit $m_s<5$~keV on the neutrino mass
(this limit was, however recently revised in \cite{abazajian05}, who finds the
restriction $m_s<8$~keV).  It was also noted by~\cite{Fuller:01b,Mapelli:05}
that one can obtain even stronger constraints $m_s \lesssim 2$~keV -- from
diffuse extragalactic X-ray background (XRB) under the assumption, that the
dark matter in the Universe is uniformly distributed up to the distances,
corresponding to red shifts $z\ll 1$. Together with the claim
of~\cite{Hansen:01,Viel:05}, putting lower bound $m_s>2$~keV on the neutrino
mass from Lyman $\alpha$-forest observations, this would lead to a very narrow
window of allowed sterile neutrino masses, if not exclude it completely.

In this paper we re-analyze the limit imposed on the parameters of sterile
neutrino by the observations of the diffuse X-ray background (XRB). For that
we are processing actual astrophysical data of HEAO-1 and XMM-Newton missions.
There are several motivations for this. Namely
\begin{enumerate} 
\item All the above restrictions on sterile neutrino mass
  (\cite{Fuller:01b,Dolgov:02,Mapelli:05,abazajian05}) are model dependent and
  based on the assumption that sterile neutrinos were absent in the early
  Universe at temperatures larger than few GeV.  Depending on the model, the
  relation between the mass of sterile neutrino, the mixing angle and the
  present-day sterile neutrino density $\Omega_s$ does change. In fact, to
  compute the sterile neutrino abundance one needs to know whether there is
  any substantial lepton asymmetry of the universe at the time of sterile
  neutrino production, what is the coupling of sterile neutrino to other
  particles such as inflaton or super-symmetric particles, etc.\footnote{For
    example, if the coupling of sterile neutrino to inflaton is large enough,
    the main production mechanism will be the creation of sterile neutrinos in
    inflaton oscillations rather than active-sterile neutrino transition.}
  Moreover, even if these uncertainties were removed, the reliable computation
  of the relic abundance of sterile neutrinos happens to be very difficult as
  the peak of their production falls on the QCD epoch of the universe
  evolution, corresponding to the temperature $\sim 150$ MeV
  (\cite{Dodelson:93}), where neither quark-gluon nor hadronic description of
  the plasma is possible. Therefore, before the particle physics model is
  fully specified and the physics of hadronic plasma is fully understood, one
  can not put a robust restriction on one single parameter of the model such
  as~$m_s$.
  
  Therefore we aim in this paper at clear separation between the model
  independent predictions, based solely on astrophysical observations and any
  statements that depend on a given model and underlying assumptions. To this
  end we treat $m_s$ and $\sin\theta$ as two \emph{independent} parameters and
  present the limits in the form of an ``exclusion plot'' in the $(m_s;
  \Omega_s \sin^22\theta)$ parameter space.
  
  It should be stressed that our data analysis is not based on any specific
  model of sterile neutrinos and as such can be applied to any ``warm'' DM
  candidate particle which has a radiative decay channel. In case of sterile
  neutrino the full decay width of this process is related to parameter
  $\sin\theta$ via Eq.~(\ref{eq:4}) (see below).
  
\item Contrary to the previous works~(\cite{Fuller:01b,Mapelli:05}), we argue
  that non-isotropy or ``clumpyness'' of the matter distribution in the nearby
  Universe \emph{does not} relax the limit on the neutrino mass.  Indeed, the
  fact that significant part of the dark matter at red shifts $z\lesssim10$ is
  concentrated in galaxies and clusters of galaxies just means that the
  strongest signal from the dark matter decay should come from the sum of the
  signals from the compact sources at $z\lesssim10$. Taking into account that
  DM decay signal from $z\lesssim10$ is some 2 orders of magnitude stronger
  than that of from $z\gtrsim10$, while the subtraction of resolved sources
  reduces the residual X-ray background maximum by a factor of 10, we argue
  that it would be wrong to subtract the contribution from the resolved
  sources from the XRB observations when looking for the DM decay signal. The
  form of XRB background spectrum with sources subtracted is, in fact, unknown
  and the assumption that it has a shape of initial spectrum, scaled down
  according to resolved fraction (as in \cite{Fuller:01b}) requires additional
  justification.
  
\item We find that more elaborate analysis of the data enables to put tighter
  limits on the allowed region of the parameter space $(m_s,\Omega_s\sin^2
  2\theta)$ from the XRB observations. The idea is that the cosmological DM
  decay spectrum is characterized not only by the total flux but also by a
  characteristic shape. 
  Being present in the XRB spectrum, it would produce a local feature with
  some clear maximum and a width greater than spectral resolution of the
  instrument. Features of such a scale are clearly absent in the data (tinier
  features could be present in the spectrum due to e.g. element lines, but
  they can not produce a bump wider than spectral resolution). Therefore,
  one can find that adding to the standard broad continuum model of XRB the DM
  decay component in a wrong place results in decrease of the overall quality
  of the fit of the data by such a two-component model (increase of the
  $\chi^2$ of the fit).  The condition that the two-component model provides
  an acceptable fit to the data imposes an upper limit on the flux in the DM
  decay component which is much more restrictive than the limit following from
  the condition that the flux of the DM component should not exceed the flux
  in the continuum component.
\end{enumerate}

The paper is organized as follows. In Section~\ref{sec:dm-flux} we compute the
contribution of the radiative decay of sterile neutrino to the diffuse X-ray
background (XRB) and compare its shape with that of measured XRB. We discuss
the effects of non-uniformness of the DM distribution in
Section~\ref{sec:uniform}. In Section~\ref{sec:diffuse-bg} we obtain a
model-independent exclusion region from HEAO-1 and XMM-Newton observations.

\section{The contribution of DM decays into the XRB.}
\label{sec:dm-flux}

The sterile neutrino with a decay width $\Gamma$ and mass $m_s$ decays into an
active neutrino and emits photons with a line-like spectrum at the energy
$E_\gamma= m_s/2$.  However, photons emitted at different cosmological
distances are redshifted on their way to the Earth so that as a result the
photon spectrum is given by (see \cite{Masso:99,Fuller:01b}):
\begin{equation}
  \label{eq:1} 
\frac{d^2 N}{d\Omega \,d E} = \frac{\Gamma
  n_{DM}^0}{4\pi} \frac1{E H\bigl(m_s/(2E)-1\bigr)}~.
\end{equation}
Here $n^0_{DM}$ is the DM number density at the present time and $H(z)$ is the
Hubble parameter as a function of red shift. The explicit form of $H(z)$
depends on the cosmological parameters. This means that the expected dark
matter decay spectrum is different for different cosmologies.  We will be
interested in $z$ corresponding to the epoch more recent than
radiation-dominated universe and, therefore, $H(z)$ can be written as
\begin{equation}
  \label{eq:2}
  H(z) \simeq H_0\sqrt{\Omega_\Lambda + \Omega_{matter}(1+z)^3},
\end{equation}
where $H_0$ is the present-day value of the Hubble parameter and
$\Omega_\Lambda,\Omega_{matter}$ are the cosmological constant and
matter contributions to the density of the Universe. Substituting
(\ref{eq:2}) into (\ref{eq:1}) we find that for small $z$ the
spectrum is approximated as
\begin{equation}
  \label{eq:3} \frac{d^2 N}{d\Omega \,d E} \simeq\frac{\Gamma
  n_{DM}^0}{2\pi H_0}
  \frac{(2E)^{1/2}}{\sqrt{8E^3\Omega_\Lambda+\Omega_{matter}m_s^3} }~.
\end{equation}
Assuming the Majorana nature of neutrino mass, the decay width $\Gamma$ is
related to $\sin2\theta$ via~\citep{Pal:81,Barger:95}:
\begin{equation}
  \label{eq:4}
  \Gamma = \frac{9\,\alpha\, G_F^2}{256\cdot 4\pi^4} \sin^2(2\theta)\, m_s^5 =
  5.6\times10^{-22}\sin^2\theta 
  \,\left(\frac{m_s}{1\:\mathrm{keV}}\right)^5\;\mathrm{sec}^{-1}~.
\end{equation}

\begin{figure}
  \includegraphics[width=\columnwidth]{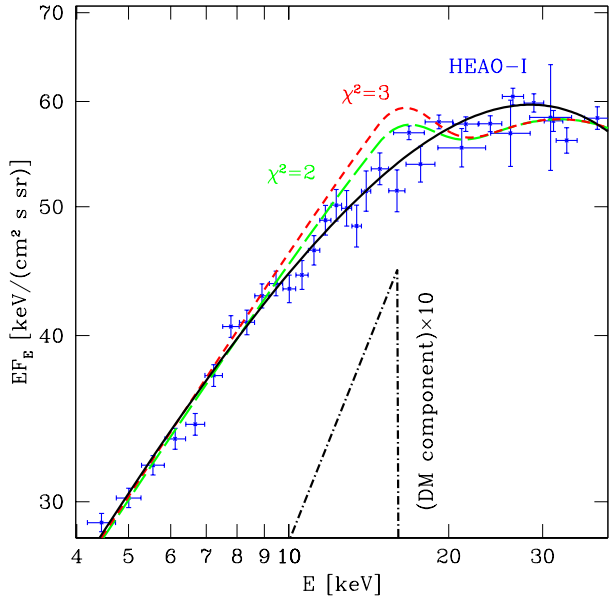} %
  \caption{Diffuse X-ray background spectrum from HEAO-1 mission. The black
    solid curve shows empirical fit Eq.~(\ref{eq:5}) by~\protect\cite{Gr99}.
    The reduced $\chi^2$ of this fit is $1.2$.  Dashed (green and red) lines
    represent the result of fit of the same data to the model of the
    form~(\ref{eq:5}) with added DM component.  The DM decay component
    (dot-dashed curve) is calculated for the ``concordance'' model
    $\Omega_\Lambda=0.7,\Omega_{matter}=0.3$ and for $m_s = 36.5$~keV.  The
    green (long dashed) line represents a fit of DM with the mixing angle
    $\sin^2 2\theta = 1.9\cdot 10^{-12}$ and the best achievable fit has the
    reduced $\chi^2 = 2$.  For the red (short-dashed) line the mixing angle is
    $\sin^2 2\theta = 2.4\cdot 10^{-13}$ and reduced $\chi^2 = 3$, which we
    choose as a border line of allowed quality of fit.}
  \label{fig:xrb}
\end{figure}

As an example we show on Fig.~\ref{fig:xrb} the expected dark matter decay
spectrum for the cosmological ``concordance'' model with $\Omega_\Lambda\simeq
0.7,\Omega_{matter}\simeq 0.3$. One can see that close to the maximum the
spectrum is roughly a power-law with the photon index equal to 1 ($dN/dE \sim
E^{-1}$).

The measurements of the diffuse XRB (\cite{Marshall:80,Gr99}) show that for
$3\lesssim E \lesssim 60$~keV its form is well approximated by the following
analytical expression
\begin{equation}
  \label{eq:5}
  \frac{d^2F_{XRB}}{dEd\Omega} =
  C_{XRB}\exp\Bigl(-\frac{E}{T_{XRB}}\Bigr)\left[\frac{E}{60\:\mathrm{keV}}\right]^{-\Gamma_{XRB}+1}\hskip-2em    \mathrm{\frac{keV}{keV \cdot sr\cdot
    sec \cdot cm^2}}\:.
\end{equation}
In Fig. \ref{fig:xrb} we show the HEAO-1 data points\footnote{We thank
  D.~Gruber for sharing this data with us and for many useful comments.}
together with the above analytical fit (solid black line).  Such a form of XRB
spectrum can be explained by AGN emission, under certain assumptions about AGN
populations~(\cite{worsley,Treister:05}).
Below $\sim 15$~keV XRB was measured by many X-ray missions and the background
was found to follow the power-law with the photon index $\Gamma_{XRB}\simeq
1.3-1.4$~(\cite{Gr99,xmm-bg,rxte-bg,Revn:04,Gilli:03,Barger:03}).  Analysis
of~\cite{Gr99} also finds $T_{XRB}=41.13$~keV and $C_{XRB}=7.9$.  The DM decay
component produces a harder spectrum, as one can see from Fig. \ref{fig:xrb}.

\subsection{The uniformness of the DM density  in the Universe}
\label{sec:uniform}

Most of the power in the very hard DM decay spectrum of Fig.~\ref{fig:xrb} is
emitted in the narrow energy interval close to the maximal energy
$E_{max}\simeq m_s/2$. From Eq.~(\ref{eq:3}) one can see that emission in this
energy range is produced by neutrinos decaying at the present epoch ($z\simeq
0$). For the case of the ``concordance'' model, the DM decay spectrum is
characterized by the very hard inverted power law $dN/dE \sim E^{1/2}$ below
the energy $\sim E_{max}/2$. The energy flux drops by an order of magnitude at
the energies $\sim E_{max}/4$ which correspond to the redshift $z\simeq 3$.
Thus, most of the DM decay signal is collected from the low redshifts.

It is known that the process of structure formation leads to
significant clustering of the dark matter at small redshifts. The
signal from DM decays at small $z$ is thus dominated by the sum of
contributions from the point-like sources corresponding to the large
DM concentrations, like galaxies or clusters of galaxies.

\cite{Fuller:01b} and \cite{Mapelli:05} argued that the clustering of the DM
at small redshifts makes if difficult for the instruments that measure the
diffuse X-ray background to detect the signal from the DM decays at small
redshifts. Indeed, measurements of the diffuse X-ray background done with
narrow-field instruments, like Chandra or XMM-Newton can miss the DM signal
because of the absence of large nearby galaxies or galaxy clusters in the
fields used for the deep observations and background measurements.  As a
measure for galaxy clustering one can take the distribution of the number of
galaxies as a function of the distance $\langle N_{\mathrm{gal}}(r)\rangle$.
The Sloan Digital Sky Survey data shows that the function $\langle
N_{\mathrm{gal}}(r)\rangle$ becomes constant for $r\gtrsim 100$~Mpc (which
corresponds to red shift $z\sim 0.02$) (see e.g.~\cite{pietronero}).
Therefore, an instrument with the field of view (FoV), which encompasses the
volume $\sim(100\;\mathrm{Mpc})^3$ at distances, corresponding to $z\lesssim
1$, will observe homogeneous matter distribution. Let us take Hubble distance
as an estimate for such distances: $H_0^{-1}\sim 3.8\times 10^3$~Mpc (here
$H_0$ is a present day Hubble constant). Then the minimal FoV $\theta_{fov}$
is determined from a simple relation $\theta_{fov}^2 H_0^{-3} \gtrsim
(100\;\mathrm{Mpc})^3$, i.e.  $\theta_{fov} \gtrsim 15'$. The FoV of
XMM-Newton is precisely of this order $\theta_{xmm}\sim 30'$. Therefore, the
strongest signal from small $z$ can be missed if only ``empty fields'' are
selected for the XMM XRB observations.


However, wide field-of-view instruments or the instruments which have
conducted all-sky surveys, like HEAO-1 or ROSAT cannot miss the largest
contribution to the DM decay signal from $z<10$ because of the full sky
coverage. Thus, the above argument should be used in a reverse sense: to find
the DM decay signal it is necessary to use the data on the X-ray background
collected from the whole sky, rather than from the ``deep field'' observations
of a narrow-field instrument.

\section{Restrictions on parameters of sterile neutrinos from the measurements
  of the diffuse X-ray background}
\label{sec:diffuse-bg}

\begin{figure}
  \includegraphics[width=\columnwidth]{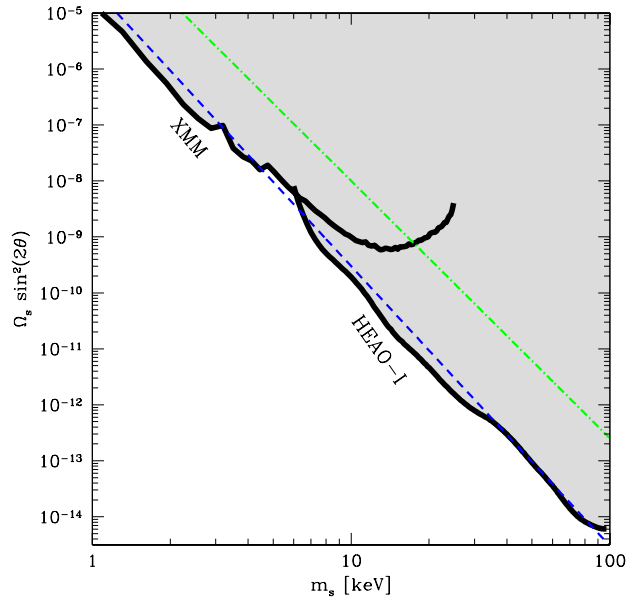} %
  \caption{Exclusion plot on parameters     $m_s$ and $\sin^2 2\theta$ using
    HEAO-1 and XMM data. The values in the non-shaded region are allowed.  In
    the region where both HEAO-1 and XMM-Newton data is available, HEAO-1
    provides a more stringent constraint (as discussed in
    Section~\ref{sec:xmm}). We supposed that sterile neutrino constitute 100\%
    of all the DM. (i.e. $\Omega_s = \Omega_{DM}$). To remove dependence on
    the value of $\Omega_{DM}$, we choose to plot $\Omega_s \sin^2 2\theta$
    rather than $\sin^2 2\theta$. The results can be described by a simple
    empirical formula Eq.(\ref{eq:6}) (thin blue dashed line on the Figure).
    The dash-dotted green line represent exclusion region, obtained if one
    attributes all 100\% of the XRB flux to DM at energies $E\lesssim m_s/2$.}
\label{fig:exclusionFromHEAO}
\end{figure}

\subsection{Restriction from HEAO-1 measurements}
\label{sec:heao}

The above analytical approximation (\ref{eq:5}) provides a good fit to the
HEAO-1 data. No DM decay feature with the spectrum of the form of (\ref{eq:3})
(corrected for the spectral resolution of the instrument) is evident in the
data. Straightforward constraint on the possible contribution of DM decays
into the XRB spectrum is that the flux of the DM decay contribution does not
exceed the total flux.  Such an approach was used by \cite{Dolgov:02}, who
applied it to the broad-band XRB model of~\cite{turner}. Using the
fit~(\ref{eq:5}) of \cite{Gr99}, which gives a much better description in the
keV region, we obtain an exclusion curve, shown as green dash-dotted line on
Fig.~\ref{fig:exclusionFromHEAO}.

Clearly this restriction is model independent and rigorous. However, it can be
made stronger. Indeed, the experimentally measured XRB background is monotonic
at energies from keV to GeV (see~\cite{Gr99}).  If the
flux of the DM have composed significant part of all XRB, 
contributions to XRB, \emph{unrelated} to the radiative DM decay, should have
combined themselves into a spectrum with a sudden narrow ``dip''. In
particular, at energies $E\ge m_s/2$ some new physical phenomena should have
suddenly ``kicked in'', causing the spectrum to experience a very sharp
``jump''. 
Moreover, the form of this dip must have been almost the same as the dark
matter contribution with a minus sign.  This would have required a mechanism
of a very precise fine-tuning of contributions to XRB between DM and various
physical phenomena or simply a chance coincidence. Of course, such a
conspiracy cannot be absolutely excluded, in particular because of the lack of
unambiguous theoretical predictions of the shape of XRB spectrum.  However, it
is very unlikely to have a precise cancellation of contributions of different
nature. So, in what follows we will assume that such a situation does not
happen (in particular, this does not happen in models, attributing existing
XRB shape to AGN emissions (\cite{worsley,Treister:05})).


Therefore, in the present work we argue that the constraint on the possible DM
contribution to the XRB resulting from HEAO-1 data can be improved using the
statistical analysis of the data.  Our strategy will be the following. We take
the actual HEAO-1 data and fit it to the model of the form~(\ref{eq:5})
(varying $C_{XRB},T_{XRB},\Gamma_{XRB}$) plus an additional DM
flux~(\ref{eq:3}) (corrected for the spectral resolution of HEAO-1
instrument). Addition of a large DM contribution would worsen the quality of
the model fit to the data, while addition of a small DM contribution does not
change the overall fit quality. Thus, one can put a restriction on the DM
contribution by allowing the DM component to worsen the fit by a certain
value. Taking into account that we have around 40 degrees of freedom of the
system under consideration, we take the maximal allowed value of the reduced
chi square of the fit to be $\chi^2 < 3$. Thus on technical level our method
restricts the flux of DM to be of the order of errors of measured XRB flux,
rather than its total value.

The results of the application of the above algorithm to the data are shown on
the Fig.~\ref{fig:exclusionFromHEAO}.  The allowed values of $({m_s},
\Omega_s\sin^2 2\theta)$ are those in the unshaded region.\footnote{The
  normalization of XRB spectrum, measured by HEAO-1, is known to be lower than
  any other XRB measurements (\cite{Gilli:03,Moretti:03}). Therefore on
  Fig.~\ref{fig:exclusionFromHEAO} HEAO-1 data were \emph{increased} by 40\%
  according to~\cite{DeLuca:03,worsley,Treister:05} (see,
  however~\cite{Gilli:03}).  This weakens the restriction of
  Fig.~\ref{fig:exclusionFromHEAO} by about 10\%, as compared to actual HEAO-1
  data.}
The shaded region below the dash-dotted line is excluded under the assumption
that XRB spectrum does not have an unknown feature (a ``dip''), fine-tuned to
be located precisely at energies where DM contributes.

According to Eqs.~(\ref{eq:3})--(\ref{eq:4}) flux of DM is proportional to
$\Omega_s \sin^2 2\theta$. Therefore we choose to plot this value rather than
conventional $\sin^2 2\theta$ on $y$ axis of Fig.~\ref{fig:exclusionFromHEAO}.
This removes uncertainty, related to the determination of $\Omega_{DM}$.
Notice, that the results obtained from HEAO-1 data (thick solid lines on
Fig.~\ref{fig:exclusionFromHEAO}) are model independent and can be applied to
any DM candidate, which has a decay channel into a lighter particle and a
photon, with its decay width $\Gamma$ related to $\sin\theta$ via
Eq.~(\ref{eq:4}).

\subsection{Restriction from XMM background measurements}
\label{sec:xmm}

In the energy band below 10~keV the XRB was studied by numerous narrow FoV
instruments, including XMM-Newton. Better angular resolution and sensitivity
of these instruments has enabled to resolve some 90\% of the XRB into point
sources~(\cite{chandra-deep,DeLuca:03,Bauer:04}). Apparently, better
sensitivity should enable to put tighter constraints on the possible DM
contribution into the XRB. However, in this section we show that this in not
the case. 

The key point which leads to such a conclusion is that the better sensitivity
(to point sources) of these instruments is achieved due to the better angular
resolution, rather than due to the increase of the effective collection area.
Thus, if one is interested in the diffuse sources, the sensitivity is not
improved compared to HEAO-1. On the contrary, in the narrow FoV instruments
the XRB signal is collected from a smaller portion of the sky, which leads to
lower statistics of the signal. At the same time, the instrumental background
(which is thought to be due to the cosmic rays hitting the instrument) is
roughly the same for narrow and wide field instruments. Thus, the ratio of the
counts due to the photons of the XRB to the instrumental background counts is
smaller for the narrow field instruments and one needs larger integration
times to achieve good statistical significance of the XRB signal.

The above problem would not have affected the constraints on $m_s$,
$\sin\theta$ if they were imposed by the condition for the flux in the DM
decay component not to exceed the total XRB flux in a given energy interval.
However, as the statistical significance of the XRB signal in the narrow FoV
instruments is lower, bigger errors decrease $\chi^2$ value, and therefore the
imposed limit on DM will be worse than that of the corresponding wide FoV
instruments.

The exclusion plot in the $(m_s,\Omega_s\sin^22\theta)$ parameter space
obtained from the analysis of the XMM data, is shown in
Fig.~\ref{fig:exclusionFromHEAO}. We took the actual data of two collections
of XMM background observations (\cite{xmm-bg}, total exposure time $\sim
450$~ksec and \cite{birm-bg} total observation time $\sim 1$~Msec). The form
of the XMM background is fitted by the power-law with the index $\Gamma=1.4$
(\cite{xmm-bg}), in agreement with Eq.~(\ref{eq:5}).  Applying to XMM data our
method, as described in the previous Section, we find the restrictions, shown
on Fig.~\ref{fig:exclusionFromHEAO}.  In the neutrino mass region
$6~\mathrm{keV}\lesssim m_s \lesssim 20$~keV where both HEAO-1 and XMM data
are available, the XMM-Newton background provides weaker restriction that the
data from HEAO-1, because of the reason discussed above (the XMM FoV has the
radius 15', which is much smaller than that of the HEAO-1, $\sim 3^\circ$ for
A2 HED detectors).  In case of XMM the statistical error at energies above
roughly 7~keV is dominated by instrumental background, while in case of
HEAO-1, the errors at these energies were dominated by diffuse X-ray
background. As a result, errors of flux determination are smaller for HEAO
data, thus providing more stringent restrictions.  Of course, the XMM data
provides the constraints on parameters of the neutrino in the region
$1\lesssim m_s\lesssim 6$ keV where HEAO-1 data are not available.  The data
on Fig.~\ref{fig:exclusionFromHEAO} fits to the simple empirical formula:
\begin{equation}
  \label{eq:6}
  \Omega_s\sin^2(2\theta)<3\times 10^{-5}\left(\frac{m_s}{\mathrm{keV}}\right)^{-5}\:.
\end{equation}

\subsection{Discussion.}
\label{sec:numsm}

In this work we have looked for signatures of sterile neutrino decay in the
extragalactic X-ray background. The main result of our paper is the plot on
Fig.~\ref{fig:exclusionFromHEAO} which constraints the properties of sterile
neutrino as a dark matter candidate.

Let us compare our exclusion plot with the similar plots, found in other
works.  \cite{Dolgov:02} used a broad-band limit, put on XRB by \cite{turner},
to find a restrictions on sterile neutrino parameters. First of all, in the
keV energy band the limit of \cite{turner} is weaker than that of \cite{Gr99},
which we are using in this paper. Therefore the dashed-dotted line on the
Fig.~\ref{fig:exclusionFromHEAO} provides model-independent restrictions on
parameters of sterile neutrino that are stronger than the similar bound on
Fig.~4 in \cite{Dolgov:02}.  Second, as discussed in
Section~\ref{sec:diffuse-bg}, we put a restriction on parameters of sterile
neutrino, based on the statistical analysis of the actual experimental data of
HEAO-1 and XMM-Newton missions.  Such a constraint is possible under the
assumption that there is no extremely unlikely fine-tuning between various
components, contributing to the XRB.  To put it differently, the observed XRB
spectrum is monotonic in keV range (at energies below $\sim 15$ keV it was
checked by various X-ray missions, including recent measurements by XMM and
Chandra). Therefore, we assume that XRB spectrum \emph{without} DM has no
``dip'' with the shape and location precisely at the place, where DM
contributes. The results of our analysis are shown in thick black solid line
on Fig.~\ref{fig:exclusionFromHEAO}. As the statistical quality of the data is
good, this puts about two orders of magnitude stronger restrictions then
those, represented by the dashed-dotted line.  Region between the solid (or
its empirical fit, Eq.~(\ref{eq:6})) and dashed-dotted lines is excluded under
the assumptions discussed above.

Ref. \cite{Fuller:01b} provides an exclusion plot (also quoted by
\cite{abazajian05}) based on observations of the Virgo cluster.   
Ref. \cite{Boyarsky:06} shows that the restrictions from clusters of
galaxies are weaker than those of \cite{abazajian05}, however, in
mass ranges $2\kev\lesssim m_s\lesssim 10\kev$ they are 2 to 4 times
better than those, coming from XRB. For careful comparison of our
results with those of \cite{abazajian05} see~\cite{Boyarsky:06}.

In some cases, such as in the  Dodelson-Widrow (DW) scenario
(\cite{Dodelson:93})\footnote{In this scenario it is assumed that
there is just one species of sterile neutrinos, that their
concentration was zero at temperatures higher than $1$ GeV, that
there were no substantial lepton asymmetries, and that the universe
reheating temperature was larger than $1$ GeV.} there exists a 
relation between the mass of sterile neutrino, the mixing angle and
the present-day sterile neutrino density $\Omega_s$
(\cite{Dolgov:02,Fuller:01b,abazajian05}). Since both the sterile
neutrino abundance and the gamma-ray flux depend on the mixing angle
$\theta$ and sterile neutrino mass $m_s$ only, this provides an upper
limit on sterile neutrino mass in the DW scenario. According to the
computation of sterile neutrino abundance made in \cite{abazajian05},
the corresponding limit following from the X-ray bound found in the
present paper reads $m_s < 8.9$ keV for the DW sterile neutrino.
However, even in DW scenario this number is subject to uncertainties
which are difficult to estimate, since the peak of the sterile
neutrino production falls into the range of temperatures where
neither hadronic nor quark description of the hot plasma is possible.

The Standard Model augmented by just one sterile neutrino  cannot
explain the neutrino oscillations data and thus extra ingredients
should be added to it. The analysis of~\cite{Asaka:06a} shows that in
a more realistic model, the $\nu$MSM, a relation between the mass of
sterile neutrino, the mixing angle and the present-day sterile
neutrino density $\Omega_s$ changes, depending on unknown parameters,
such as the masses and couplings of extra particles and primordial
abundances of sterile neutrinos.  Therefore, for a general case one
cannot put a model-independent restriction on the mass of sterile
neutrino $m_s$, and the search for a narrow line coming from its
decays should be done in all energy ranges.

The results of this work show that a possible strategy to look for a feature
of DM decay in the XRB is to use data of ``all sky surveys'' rather than
``deep field observations'' and utilize instruments with the largest possible
FoV (rather than best angular/spectral resolution). An example of such an
instrument can be INTEGRAL, which currently plans an observation of XRB.

\subsection*{Acknowledgements} 
We would like to acknowledge useful discussions with K.~Abazajian, A.~Kusenko,
I.~Tkachev and R.~Rosner. This work was supported in part by the Swiss Science
Foundation and by European Research Training Network contract 005104
"ForcesUniverse".  O.R. was supported by a \emph{Marie Curie International
  Fellowship} within the $6^\mathrm{th}$ European Community Framework
Programme.


 \label{lastpage}
\end{document}